\input{aipcheck}

\documentclass[
    ,final            
  ]
  {aipproc}

\layoutstyle{8x11double}

\begin{document}

\title{Probing the Galactic cosmic ray flux with submillimeter and gamma ray data}


\author{S. Casanova}{
  address={Max Planck f\"ur Kernphysik, 69117, Heidelberg, Germany}
}

\author{S. Gabici}{
  address={Dublin Institute for Advance Physics, 31 Fitzwilliam Place,
  Dublin 2, Ireland}
}

\author{F. A. Aharonian}{
  address={Dublin Institute for Advance Physics, 31 Fitzwilliam Place,
  Dublin 2, Ireland}
  ,altaddress={Max Planck f\"ur Kernphysik, 69117, Heidelberg, Germany} 
}

\author{K.Torii}{
  address={Nagoya University, Nagoya, Japan}
}

\author{Y.Fukui}{
  address={Nagoya University, Nagoya, Japan}
}

\author{T. Onishi}{
  address={Nagoya University, Nagoya, Japan}
}

\author{H. Yamamoto}{
  address={Nagoya University, Nagoya, Japan}
}

\author{A. Kawamura}{
  address={Nagoya University, Nagoya, Japan}
}

\begin{abstract}

The study of Galactic diffuse $\gamma$ radiation combined with the knowledge of the 
distribution of the molecular hydrogen in the Galaxy 
offers a unique tool to probe the cosmic ray flux in the Galaxy. A methodology to study 
the level of the cosmic ray "sea" and to unveil target-accelerator systems in the 
Galaxy, which makes use of the data from the high resolution survey of the Galactic 
molecular clouds performed with the NANTEN telescope and of the data from $\gamma$-ray instruments, 
has been developed. Some predictions concerning the level of 
the cosmic ray "sea" and the $\gamma$-ray emission close to 
cosmic ray sources for instruments such as Fermi and Cherenkov Telescope Array are presented.

\end{abstract}

\maketitle

\section{Introduction}
 
Cosmic rays (CRs) up to at least ${10}^{15} \,  {\rm eV}$ are believed to 
be emitted by Galactic sources, such as supernova remnants, but no conclusive 
evidence of the acceleration has been found yet. A trace of 
ongoing cosmic ray acceleration is the $\gamma$-ray emission produced by these highly 
energetic particles when they scatter off the interstellar medium gas, mainly 
atomic, molecular and ionised hydrogen. In fact, from the hadronic collisions neutral pions emerge 
and in turn they decay into two $\gamma$s. For this reason $\gamma$-ray astronomy 
has always played a key role to probe the Galactic cosmic ray flux 
and to solve the longstanding question of the 
origin of cosmic rays. Whereas the atomic hydrogen is uniformly distributed in the Galaxy, the 
molecular hydrogen is usually aggregated in dense clouds, and the $\gamma$-ray emission 
from such clouds is particularly intense. A 
multi-frequency approach which combines the data from the upcoming and future $\gamma$-ray 
missions with the data from the submillimeter and milli-meter surveys of the molecular hydrogen 
is therefore crucial to probe the Galactic cosmic ray flux.  

In the following a methodology to make predictions of the level of the cosmic ray 
"sea" and to unveil target-accelerator systems in the Galaxy by using the 
data from the NANTEN survey from a region in Galactic longitude ${340}^{\rm o}<l<{350}^{\rm o}$ and 
Galactic latitude $-5^{\rm o}<b<5^{\rm o}$  will be described. In particular, one can put 
upper limits on the CR "sea" based on the observations 
of the dense environments in the Galaxy. If the flux from certain clouds
is below the predictions based on the locally measured CR density, this implies that 
the "sea level" is currently overestimated.

\section{The Nanten  survey}
Equipped with a 4 m submillimeter telescope, the NANTEN instrument 
surveyed the southern sky in the molecular hydrogen, by using the J=1-0 line of CO at 115.271 GHz (2.6 mm). 
The angular resolution of this survey is equal to 4 arc-min and the mass 
sensitivity is about 100 solar masses at the Galactic Centre at 8.5 kpc \cite{Fukui1,Fukui2,Fukui3}.  
Following \cite{Nakanishi1,Nakanishi2} 
a three dimensional map of the molecular hydrogen density is obtained 
by assuming a flat rotation curve model of the Galaxy with uniform velocity 
equal to $220 {\rm km/s}$. The factor $X= 1.4 \times {10}^{20}  \, e^{(R/11 \, {\rm kpc})} \, {\rm {cm}^{-2} {K}^{-1} {km}^{-1} s} $,  
where $R$ is distance from the Galactic Centre, 
is used to translate the CO integrated intensity into $H_2$ column density. 

\section{Molecular clouds as tracers of cosmic rays}
\subsection{Passive molecular clouds}

Cosmic rays diffuse in the Galactic magnetic fields for 
timescales of the order of ${10}^7 \, \rm{years}$ before escaping the Galaxy. 
During these timescales the particles accelerated by individual sources 
mix together, lose memory of their origin, and contribute to the bulk of Galactic cosmic rays 
known as the cosmic ray ``sea''. Because of the diffusion process, the injection spectra from individual 
sources, which according to the diffusive shock acceleration theory are expected to have a power-law slope close to -2, get softened. 
The cosmic ray flux measured close to the Earth, with its characteristic soft -2.7 spectrum, is 
usually assumed to be representative of the average cosmic ray flux throughout the Galaxy. 
However, the spectral feature of the $\gamma$-ray emission detected by HESS from the 
Galactic Centre region \cite{Aharonian:nature}, and EGRET observations of the 
Orion nebula \cite{Aharonian:2000} suggest that the CR density varies within the Galaxy. 
 
\begin{figure}
  \includegraphics[height=.22\textheight]{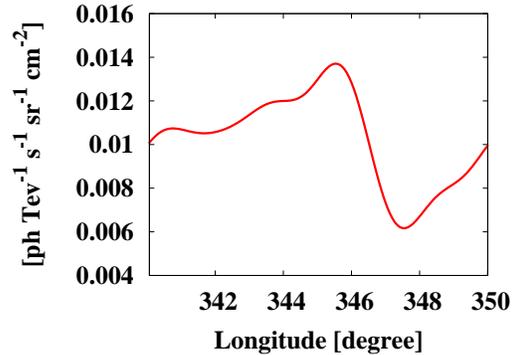}
  \caption{The longitude profile of the $\gamma$-ray emission at 1 GeV which would be measured by Fermi from the 
region $340^{\rm o}<l<350^{\rm o}$, integrated over the latitude range $-5^{\rm o}<b<5^{\rm o}$, 
if the cosmic ray flux is equal to the flux measured close to the Earth. }
\label{fig1}  
\end{figure}

\begin{figure}
  \includegraphics[height=.22\textheight]{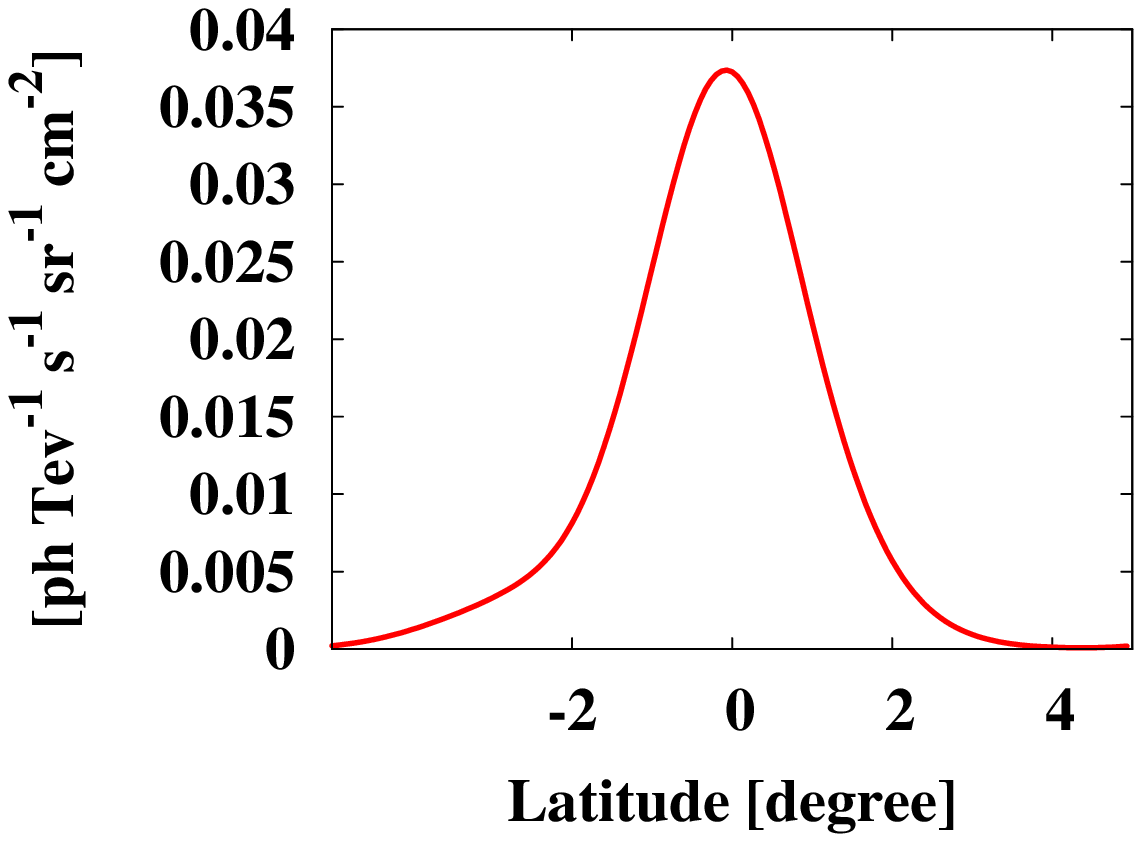}
  \caption{The latitude profile of the $\gamma$-ray emission at 1 GeV which would be measured by Fermi from the 
region $-5^{\rm o}<b<5^{\rm o}$, integrated over the longitude range $340^{\rm o}<l<350^{\rm o}$, 
if the cosmic ray flux is equal to the flux measured close to the Earth.}
\label{fig2}
\end{figure}

\begin{figure}
  \includegraphics[height=.22\textheight]{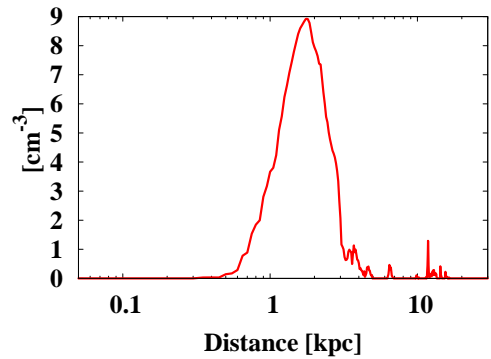}
  \caption{The gas density as a function of the line of sight distance from the direction ${346}^{\rm o}$ 
on the Galactic Plane is shown. }
\label{fig3}
\end{figure}

In order to test the assumption that the local CR flux is representative 
of the level of the cosmic ray ``sea'', we investigate the emissivity of molecular clouds located 
far away from candidates cosmic ray sources. These 
molecular clouds, illuminated by the average cosmic ray density and thus denominated ``passive'' molecular clouds, 
can be used as cosmic ray barometers. 
The detection of a single passive cloud which emits a lower $\gamma$-ray flux than that expected 
assuming the local cosmic ray flux would provide evidence that the cosmic ray flux locally 
measured is not representative of the average cosmic ray flux. 
In Figure \ref{fig1} and \ref{fig2}, respectively, the longitude and latitude profiles of the $\gamma$-ray emission 
from the Galactic region ${340}^{\rm o}<l<{350}^{\rm o}$ and  $-5^{\rm o}<b<5^{\rm o}$ due to the proton flux measured at Earth are plotted. 
There is a clear peak in the emission at about ${346}^{\rm o}$ along the Galactic Plane, next to a dip. Since the proton flux 
in the whole region is assumed to be constant, the peak in the $\gamma$ emission must be due to an enhancement
of the gas density somewhere along the line of sight from the direction ${346}^{\rm o}$ longitude and $0^{\rm o}$ latitude. 
Statistically speaking such an enhancement of the gas density from this direction 
can be due to a single or at most few very dense molecular clouds. In Figure \ref{fig3} the gas density as a function 
of the line of sight distance from the longitude ${346}^{\rm o}$ on the Galactic Plane is shown. The gas density is 
particularly high only within 0.5 and 2 kpc distance, where it amounts to about $9 \, {\rm {cm}^{-3}}$, and thus the peak 
in the $\gamma$ emission in Figure \ref{fig1} 
is mostly due to the cosmic ray flux within 0.5 and 2 kpc distance from the Sun. Notably, for such densities no question 
about the CR penetrability within the cloud arises. The $\gamma$-ray flux from 
such region, which Fermi would detect above 1 GeV is $2 \times {10}^{-8} \,{ \rm photons/({cm}^2 s)}$, and for Cherenkov Telescope Array (CTA) 
above 100 GeV is about $1 \times {10}^{-11} \, { \rm photons/({cm}^2 s)}$, which for observing a 1 degree extended region is close to 
the sensitivity of the detector. In this way the cosmic ray flux at ${346}^{\rm o}$ longitude on the Galactic 
Plane at a distance within 0.5 and 2 kpc can be estimated, and upper limits can be put on the CR "sea". If the flux from these clouds
is below the predictions based on the locally measured CR density, this implies that the "sea level" is currently overestimated \cite{Casanova2008}.

\subsection{Active molecular clouds}

Cosmic ray sources are believed to accelerate hadrons and heavier nuclei 
up to at least the so called "knee" at about ${10}^{15} \, {\rm eV}$. The direct observation of cosmic rays from 
these candidate injection sites is not possible since CRs diffuse into the Galactic magnetic fields, and the contributions
from individual sources merge into the "sea" of background cosmic rays, losing the information
on the original acceleration locations and spectra. However, before diffusing into the magnetic fields,
the newly injected cosmic rays scatter off the local gas and produce $\gamma$-ray emission through pion decay,
which can significantly differ from the the diffuse emission contributed by
the "sea" CRs because of the hardness of the young spectrum, not yet steepened by diffusion \cite{Aharonian:1996,Gabici}.
The extension of such diffuse sources does not generally exceed a few hundred parsecs,
the scale at which the spectra of freshly injected CRs can significantly differ from the spectrum of the CR "sea". 
These diffuse sources are often correlated with dense molecular clouds which
act as a target of the local enhanced CR injection spectrum. \cite{Montmerle,Casse} have pointed out 
that SNRs, the candidate CR sources {\it par excellence}, are located in star forming regions, 
which are rich in molecular hydrogen. In other words, CR sources and molecular clouds 
are associated and target-accelerator systems are not unusual within the Milky Way. 

Successful and promising as it is \cite{Aharonian:nature}, the idea of probing 
CR accelation sites with $\gamma$-ray observations is not exempt from problems.  
First, it is difficult to draw conclusions from observations because every observation 
looks different from the others. In fact, the $\gamma$-ray radiation from 
CR sources depends not only on the total power emitted by the sources in cosmic rays, and on the distance 
of the source, but also on the density of the local interstellar gas or target, on the local 
diffusion coefficient of the accelerated particles, and on the injection history of the source. 
As shown by the surveys of the Galaxy published by EGRET
at MeV-GeV energies \cite{Hartmann}, by HESS at TeV energies \cite{Aharonian:2006} 
and more recently at very high energies by the Milagro Collaboration {\cite{Abdo:2007} the 
various sources differ in spectra, flux and morphology. Secondly, most of these sources especially 
at GeV energies lack a counterpart at other frequencies, due 
to the poor angular resolution obtained by the instruments at GeV energies such as EGRET. 
Also, the source populations at GeV and TeV energies do not seem to 
coincide, with few spatially coincident and spectrally compatible sources. Again, this is 
partly due to the poor angular resolution at GeV energies, but it might also be a consequence of 
the energy dependence of the physical processes involved, such as injection and diffusion. 
In other words it is difficult to definitely recognize the sites of cosmic ray 
acceleration, since very often only qualitative predictions 
are provided, rather than robust quantitative predictions, especially from a morphological point of view. 
The only way to properly model what we expect to observe is to convey 
in a quantitave way all information by recognizing that the enviroment, the source age, 
the acceleration rate and history, all play a role in the physical process of injection and 
all have to be taken into account for the predictions. 
A first step of such an investigation consists in fruitfully taking advantage of the 
interstellar medium data provided by the molecular hydrogen surveys in order to make robust 
predictions concerning the spectral and morphological features in $\gamma$-rays we expect to observe 
from CR sources. Hereafter we have assumed that a SNR, located at 1 kpc distance, within the region 
${340}^{\rm o}<l<{350}^{\rm o}$ in longitude and  $-5^{\rm o}<b<5^{\rm o}$ in latitude, 
has happened 10,100,1000 or 10000 years ago. In particular, the location of the SNR is randomly chosen 
to be at ${342}^{\rm o}$ longitude and  $0^{\rm o}$ latitude. The gas density of the region 
surrounding the SNR event does not exceed $9 \, {\rm {cm}^{-3}}$. 
The burstlike event is assumed to have injected ${10}^{50} \,{\rm ergs}$ in cosmic rays. The 
CR injection spectra are assumed to be power-law with slope -2.2. The diffusion coefficient close 
to the SNR is ${10}^{28} \, {(\frac{E}{10 \, {\rm GeV}})}^{1/2} \, {\rm {cm}^2/s}$. 
The corresponding $\gamma$-ray spectra which Fermi and CTA would detect are shown in Figure \ref{fig4} and 
\ref{fig5}, respectively \cite{Casanova2008}. Notably the gas enviroment where the SNR is supposed to have exploded 
is not particularly dense. If the same SNR event had happened in more dense region, for example 10 times more dense, it 
would be detectable up to about 1000 years. 
\begin{figure}[htp]
\centering
  \includegraphics[height=.12\textheight]{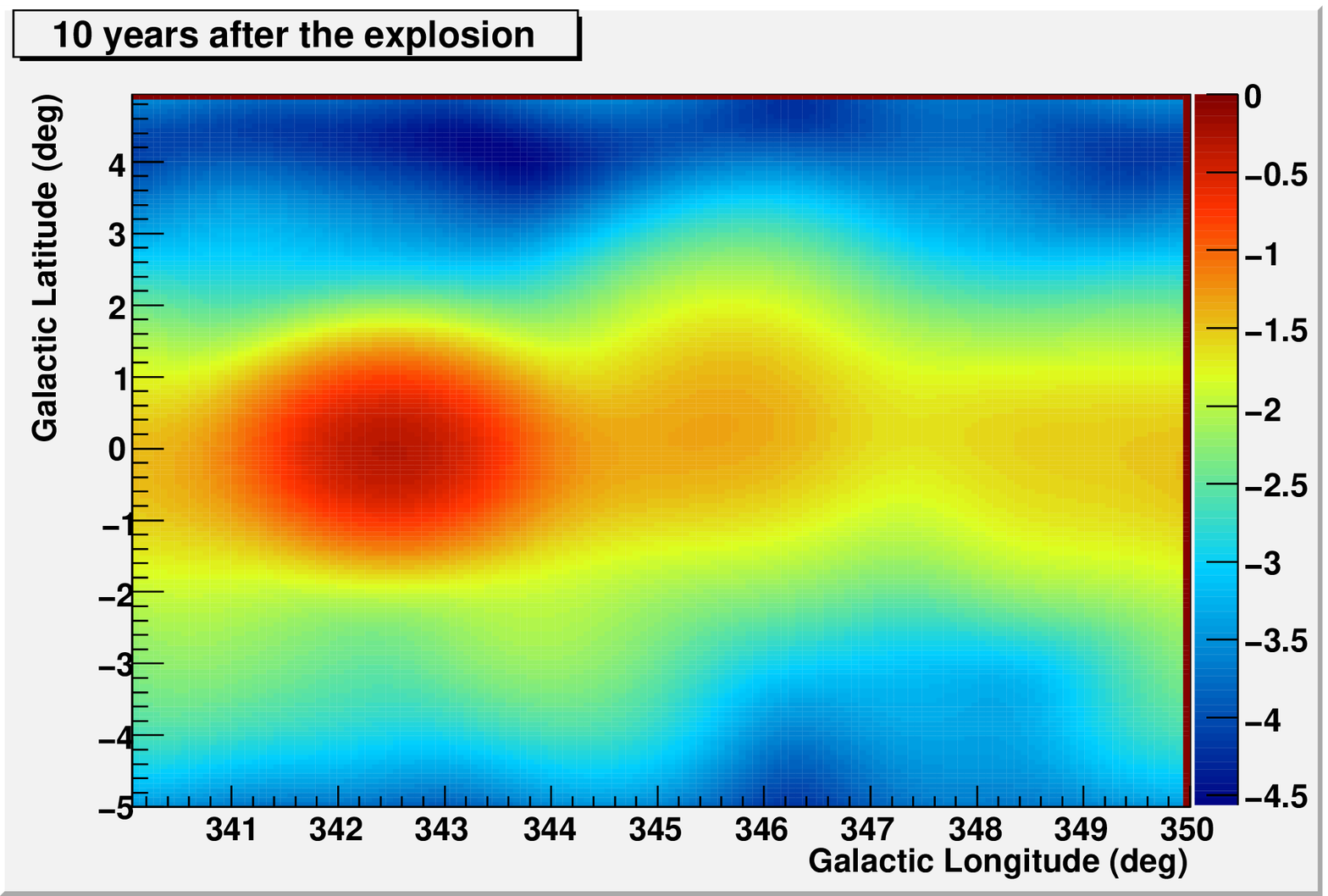}
\vspace{0.1cm}
  \includegraphics[height=.12\textheight]{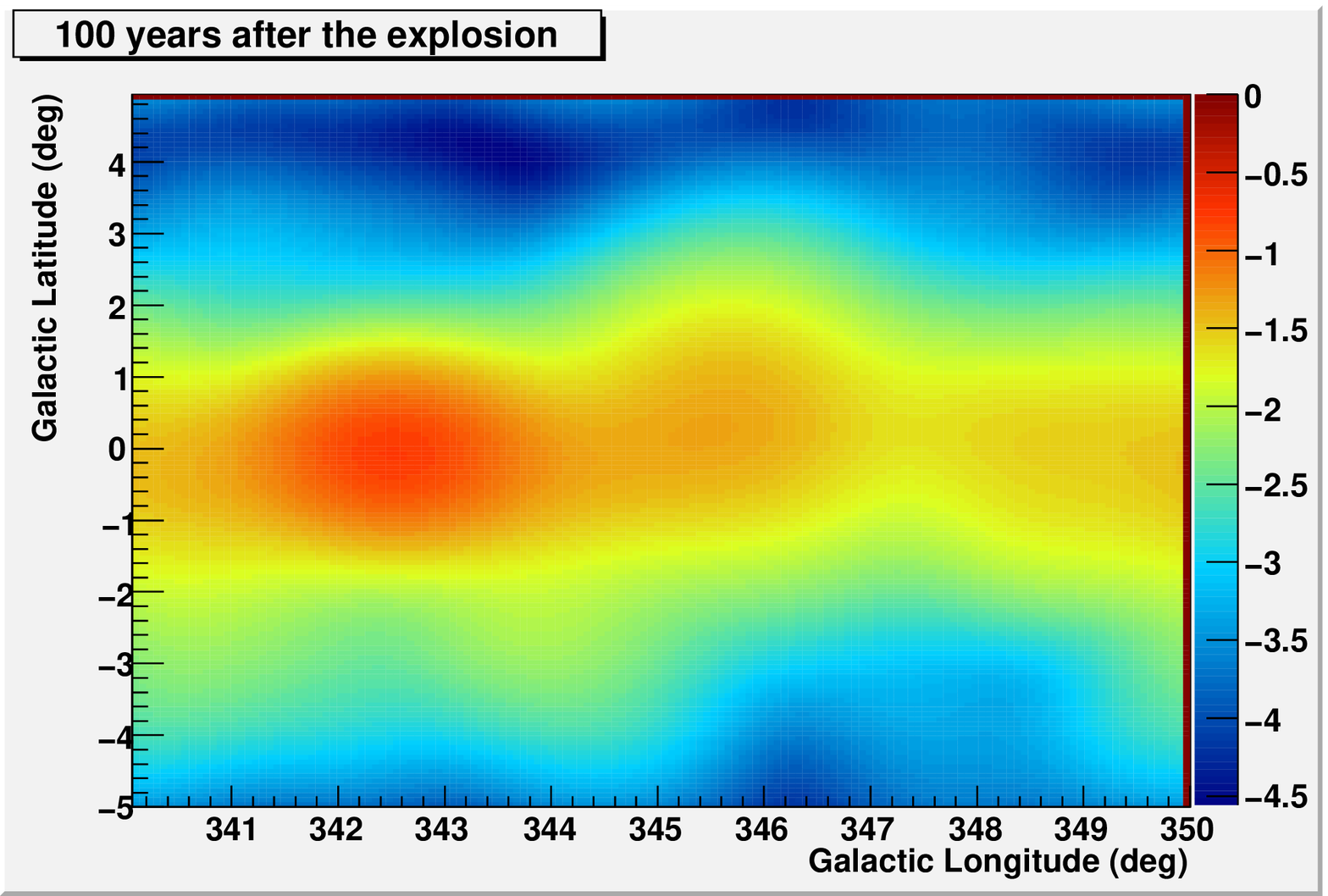}
\vspace{0.1cm}
  \includegraphics[height=.12\textheight]{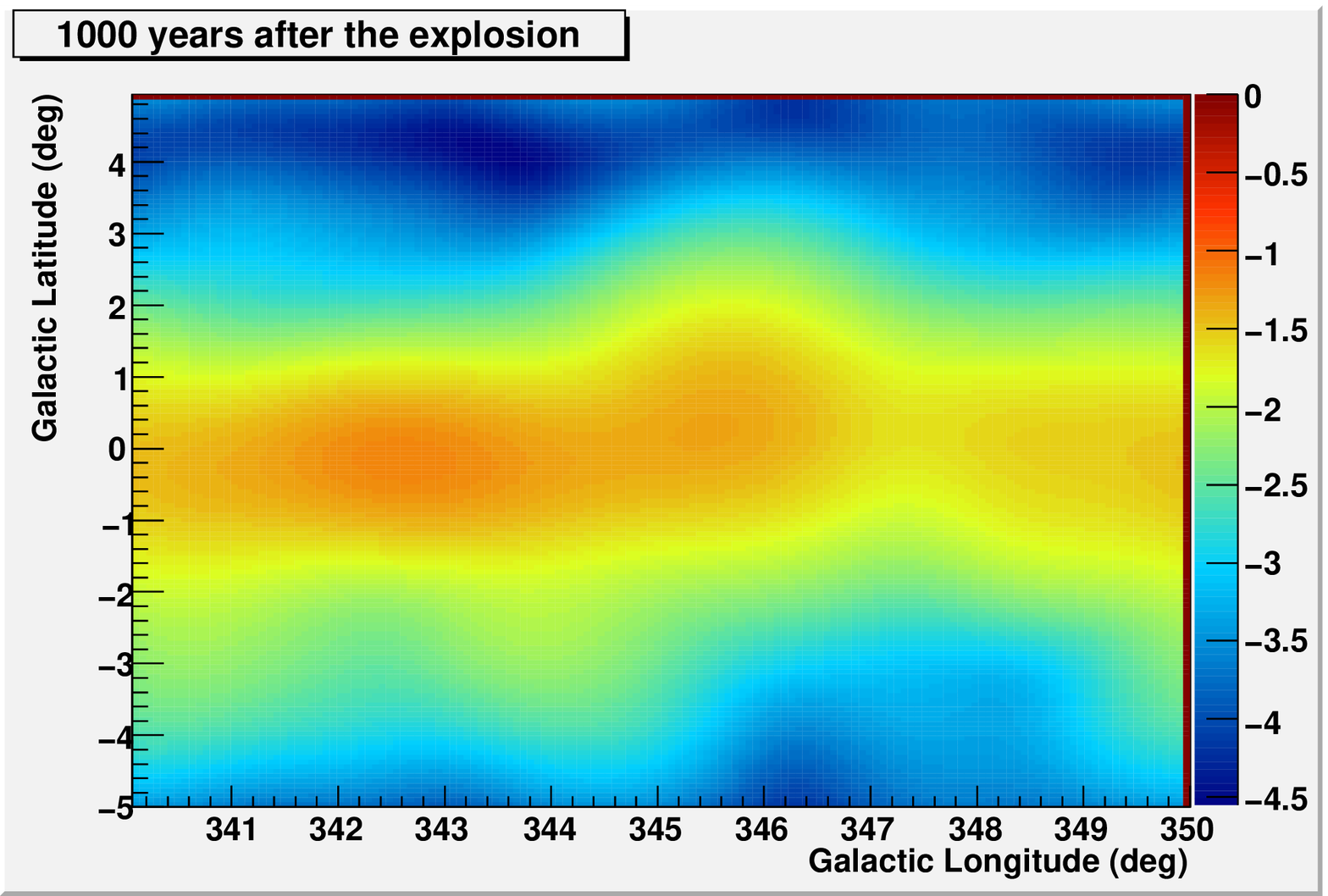}
\vspace{0.1cm}
  \includegraphics[height=.12\textheight]{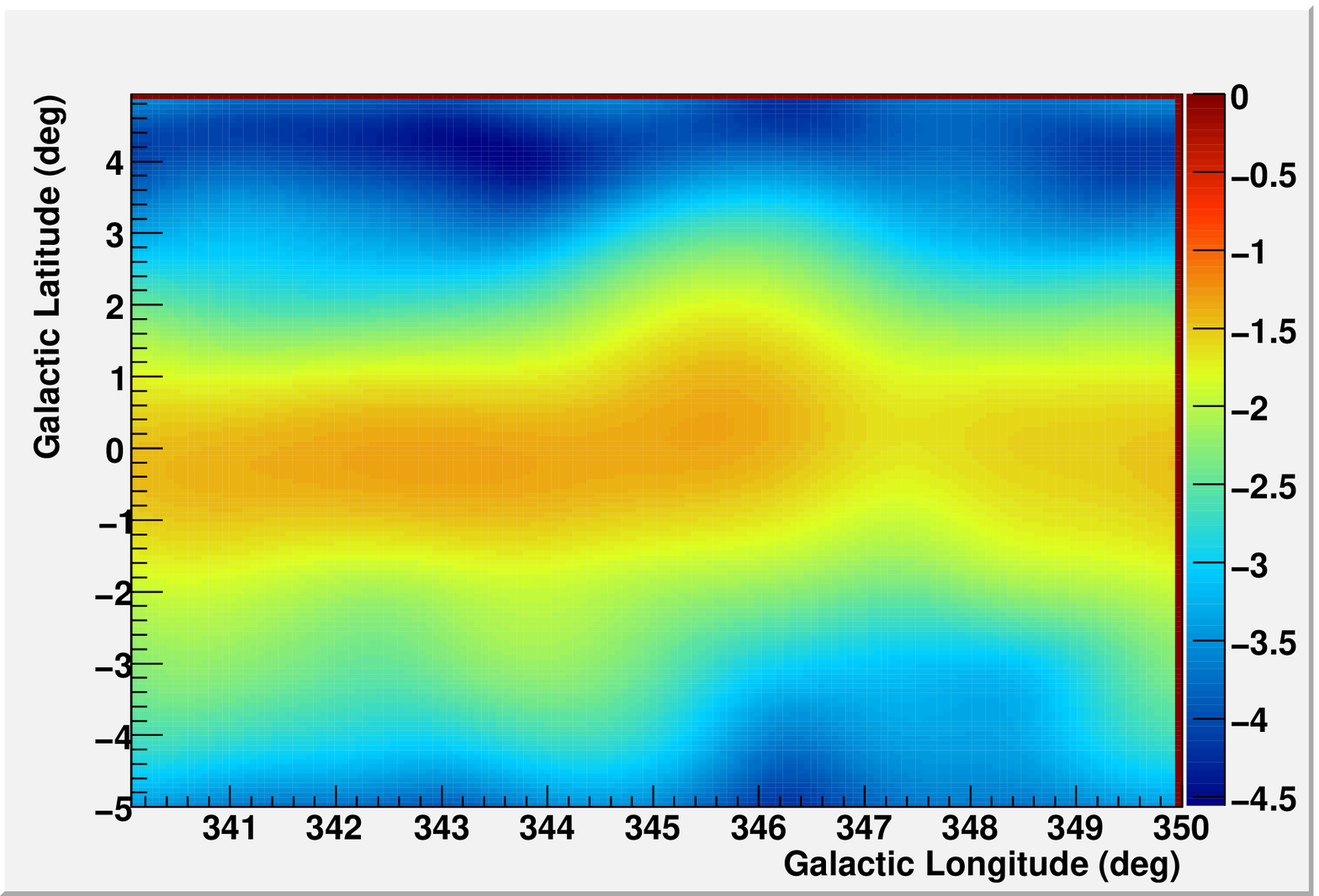}
\vspace{0.1cm}
\caption{A SNR has exploded at ${342}^{\rm o}$ longitude and $0^{\rm o}$ latitude 
10,100,1000 or 10000 years ago. The $\gamma$-ray spectrum which Fermi would detect 
at 1 GeV is expressed in ${log}_{10} \,{\rm photons /(TeV {cm}^2 sr s)}$. For reference Fermi 
point source sensitivity at 1 GeV is $2 \times {10}^{-6} \, {\rm photons/(TeV {cm}^2 s)}$. The $\gamma$-ray spectrum 
10000 years after the SNR explosion is comparable with the $\gamma$-ray spectrum produced 
by the CR sea. }\label{fig4}
\end{figure}
\begin{figure}[htp]
\centering
\includegraphics[height=.162\textwidth]{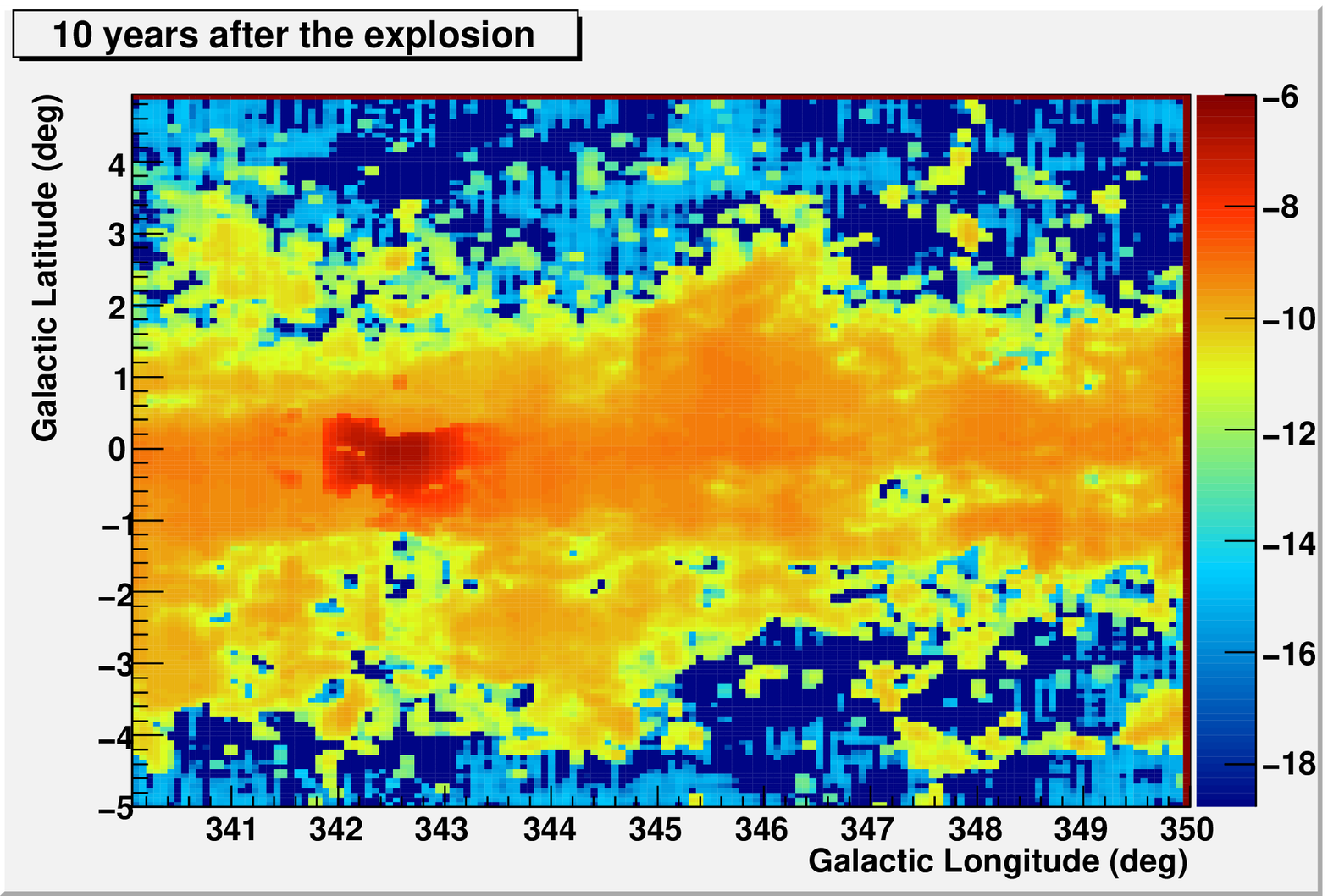}
\vspace{0.1cm}
\includegraphics[height=.162\textwidth]{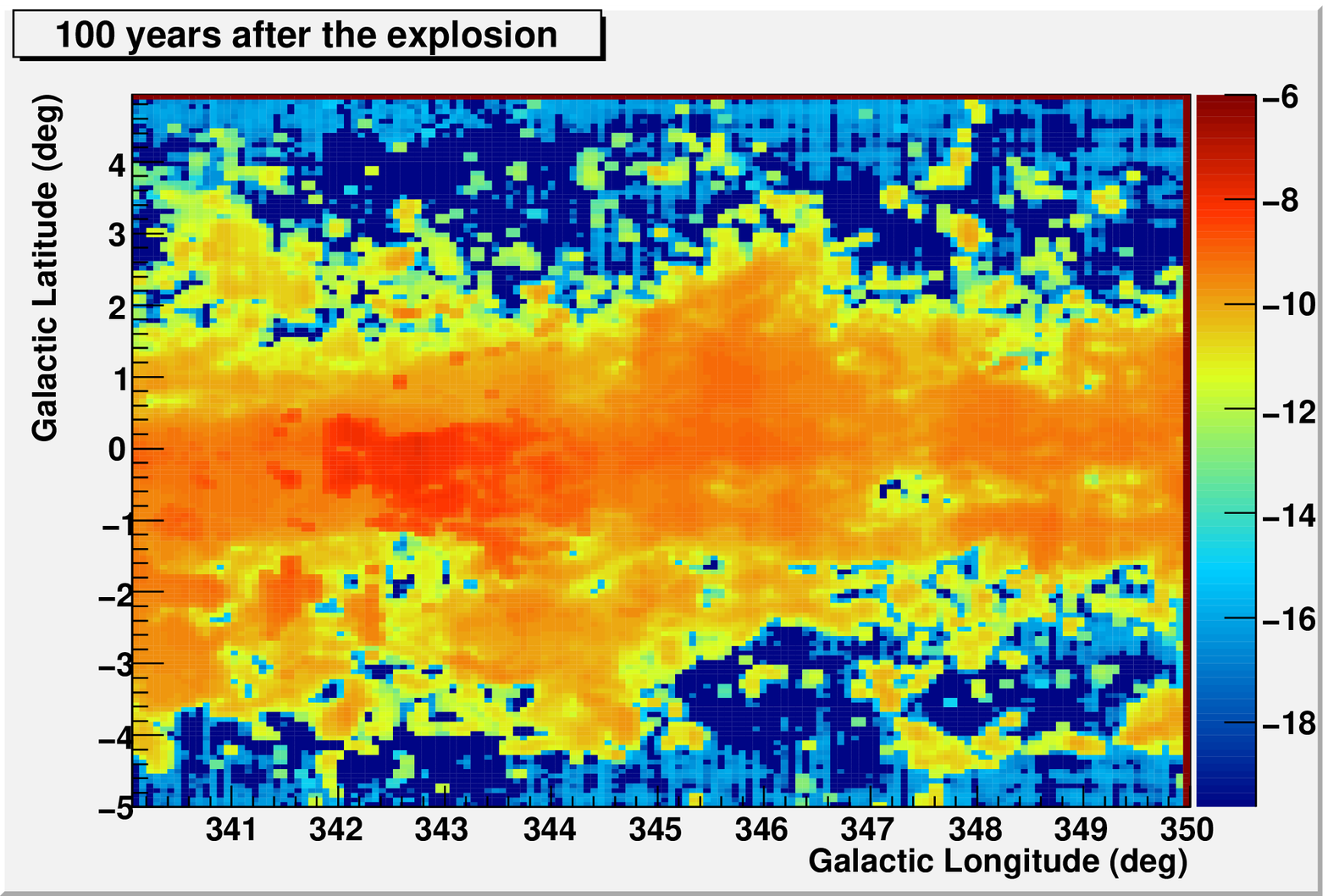}
\vspace{0.1cm}
\includegraphics[height=.162\textwidth]{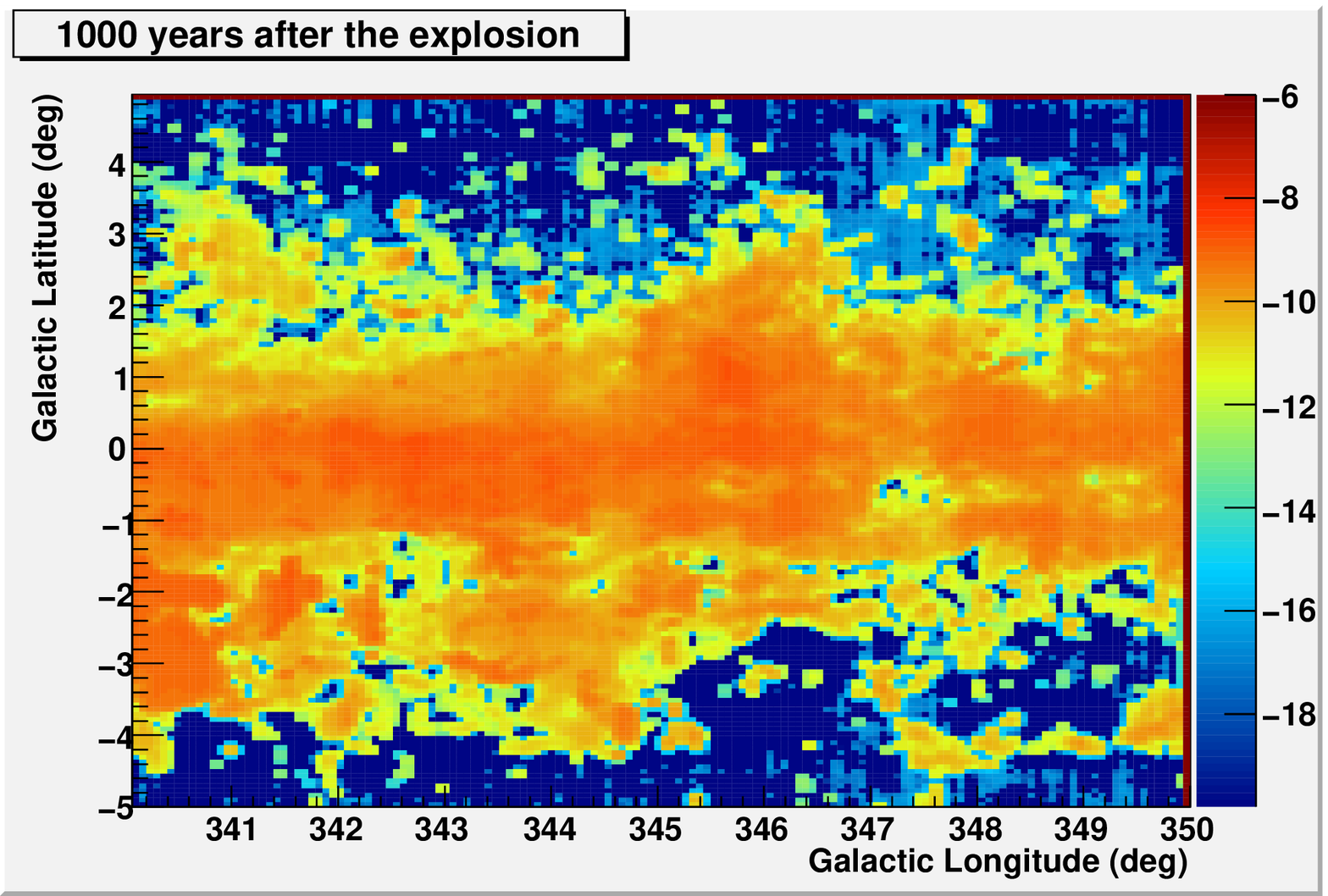}
\vspace{0.1cm}
\includegraphics[height=.162\textwidth]{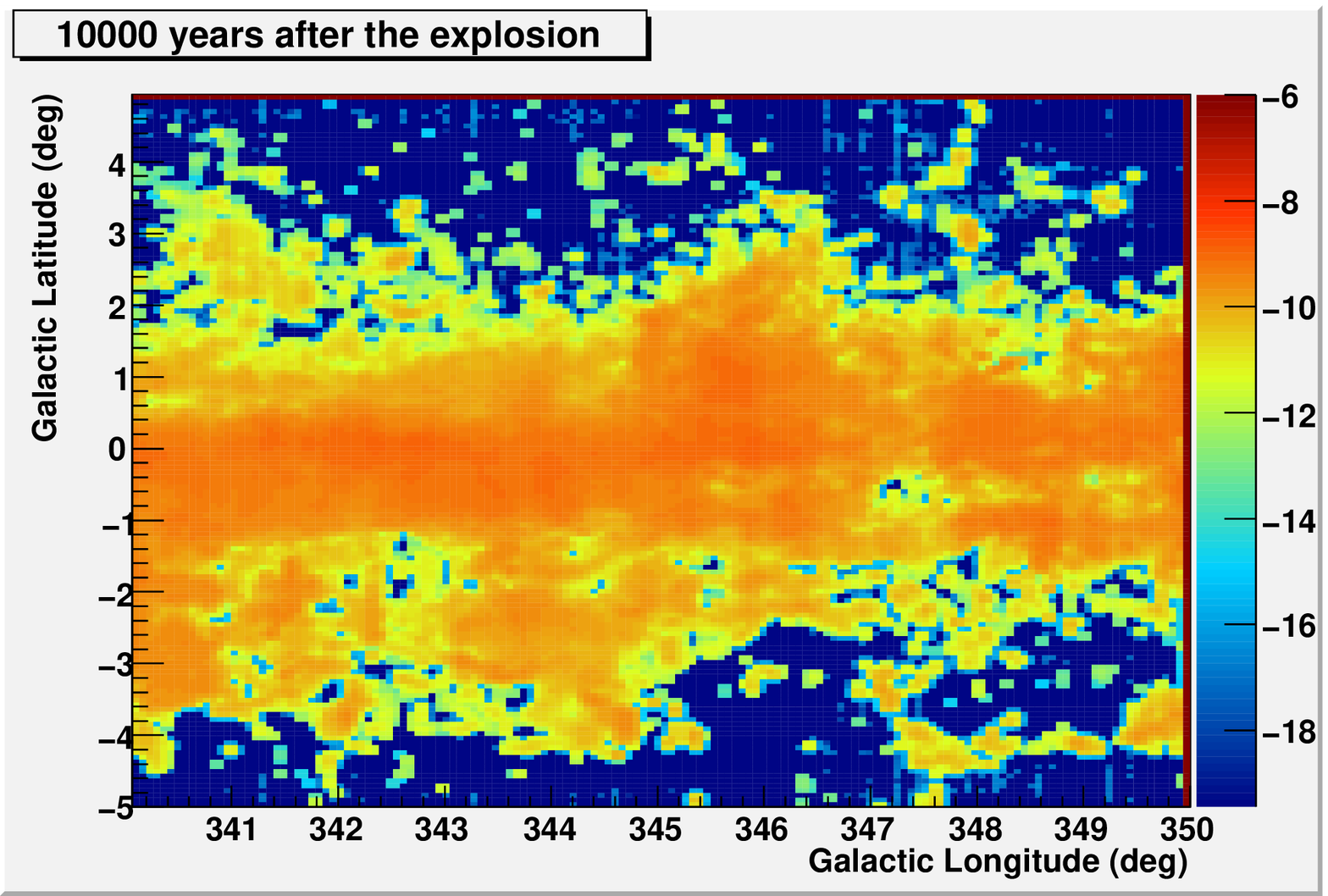}
\vspace{0.1cm}
\caption{A SNR has exploded at ${342}^{\rm o}$ longitude and $0^{\rm o}$ latitude 
10,100,1000,10000 years ago. The $\gamma$-ray spectrum which CTA would detect
at 1 TeV is expressed in ${log}_{10} \,{\rm photons /(TeV {cm}^2 sr s)}$. For reference 
CTA sensitivity at 1 TeV is about $3 \times {10}^{-14} \, (\frac{\theta}{PSF}) \, {\rm photons/(TeV {cm}^2 s)}$, 
where $\theta$ is the viewing angle and the point spread function at 1 TeV is assumed to be $PSF = {0.01}^{o}$ \cite{Aharonian2008}. 
The $\gamma$-ray spectrum 10000 years after the SNR explosion is comparable with the $\gamma$-ray spectrum produced 
by the CR sea. }\label{fig5}
\end{figure}

\section{Conclusions}
A methodology to study the level of the cosmic ray "sea" and to unveil target-accelerator systems in the 
Galaxy, which makes use of the data from the high resolution survey of the Galactic 
molecular clouds performed with the NANTEN telescope and of the data from $\gamma$-ray instruments, 
has been developed. Some predictions concerning the level of 
the cosmic ray "sea" and the $\gamma$-ray emission close to 
cosmic ray sources for instruments such as Fermi and Cherenkov Telescope Array are presented.


\begin{theacknowledgments}

The NANTEN telescope was operated based on a mutual agreement between  
Nagoya University and the Carnegie Institution of Washington. We also  
acknowledge that the operation of NANTEN was realized by contributions  
from many Japanese public donators and companies. This work is financially 
supported in part by a Grant-in-Aid for Scientific 
Research from the Ministry of Education, Culture, Sports, Science and  
Technology of Japan (Nos. 15071203 and 18026004, and core-to-core  
program No. 17004) and from JSPS (Nos. 14102003, 20244014, and  
18684003). 

Sabrina Casanova and Stefano Gabici acknowledge the support from the 
European Union under Marie Curie Intra-European fellowships. 
\end{theacknowledgments}

\bibliographystyle{aipproc}   

\end{document}